\documentclass[twocolumn,showpacs,preprintnumbers,showkeys,superscriptaddress,prc]{revtex4-1}
\usepackage{CJK}
\usepackage{mathrsfs}
\usepackage{amssymb}
\usepackage{amsmath}
\usepackage{graphicx}
\usepackage{dcolumn}
\usepackage{bm}
\usepackage{graphicx}
\usepackage{float}
\usepackage[normalem]{ulem}
\usepackage{color}
\usepackage{multirow}

\newcolumntype{d}[1]{Dc{.}{.}{#1}}
\begin{document}
\begin{CJK*}{UTF8}{} 

\title{Robust correlations between quadrupole moments of low-lying $2^+$ states within random-interaction ensembles}

\author{Y. Lei ({\CJKfamily{gbsn}雷杨})}
\email{leiyang19850228@gmail.com}
\affiliation{Key Laboratory of Neutron Physics, Institute of Nuclear Physics and Chemistry, China Academy of Engineering Physics, Mianyang 621900, China}
\date{\today}

\begin{abstract}
In the random-interaction ensembles, three proportional correlations between quadrupole moments of the first two $I^{\pi}=2^+$ states robustly emerge, including $Q(2^+_1)=\pm Q(2^+_2)$ correlations consistently with realistic nuclear survey, and the $Q(2^+_2)=-\frac{3}{7}Q(2^+_1)$ correlation, which is only observed in the $sd$-boson space. These correlations can be microscopically characterized by the rotational SU(3) symmetry and quadrupole vibrational U(5) limit, respectively, according to the Elliott model and the $sd$-boson mean-field theory. The anharmonic vibration may be another phenomenological interpretation for the $Q(2^+_1)=- Q(2^+_2)$ correlation, whose spectral evidence, however, is insufficient.
\end{abstract}
\maketitle
\end{CJK*}

\section{interaction}
Finite many-body systems (e.g., nuclei, small metallic grains, metallic clusters) robustly maintain similar regularities, despite their different binding interactions. For example, they all present the odd-even staggering on their binding energies, which are, however, attributed to various mechanisms \cite{oes-1,oes-2,oes-3,oes-4,oes-5}. Particularly in nuclear systems, the nucleon-nucleon interactions numerically exhibit a ``random" pattern with no trace of symmetry groups, whereas nuclear spectra follow some robust dynamical features: the nuclear spectral fluctuation is universally observed \cite{bohigas,haq,shriner}; low-lying spectra of even-even nuclei are orderly and systematically characterized by seniority, vibrational and rotational structures \cite{casten-1,casten-2}, beyond $I^{\pi}=0^+$ ground states without exception. 

To demonstrate the insensitivity of these robust regularities to the interaction details, and to reveal its underlying origin, random interactions are employed to simulate (or even introduce) the variety and chaos into a finite many-body system. Thus, the predominant behaviors in a random-interaction ensemble correspond to dynamical features in a realistic system. Many efforts have been devoted along this direction \cite{rand-rev-1,rand-rev-2,rand-rev-3,rand-rev-4,rand-book}. For instance, similarly to realistic even-even nuclei, the predominance of the $I=0$ ground states \cite{johnson-prl,johnson-prc} and collective band structures \cite{bijker-prl,bijker-prc} have been observed in random-interaction ensembles. However, there are only few attempts to study the robustness of nuclear quadrupole collectivity against the random interaction. This is partly because a random-interaction ensemble potentially gives weaker E2 transitions than a shell-model calculation with ``realistic" interactions \cite{horoi-be2}. Even so, some robust correlations about the E2 collectivity can be expected. For example, the Alaga ratio between the quadrupole moment ($Q$) of the $2^+_1$ state and B(E2, $2^+_1\rightarrow 0^+_1$) highlights both near-spherical shape and well deformed rotor in random-interaction ensembles \cite{rand-rev-2,horoi-a}; ratios of E2 transition rates between yrast $0^+_1$, $2^+_1$ and $4^+_1$ states are also correlated to the ratio of $2^+_1$ and $4^+_1$ excitation energies \cite{bijker-prl,zhao-be2}.

\begin{figure}
\includegraphics[angle=0,width=0.45\textwidth]{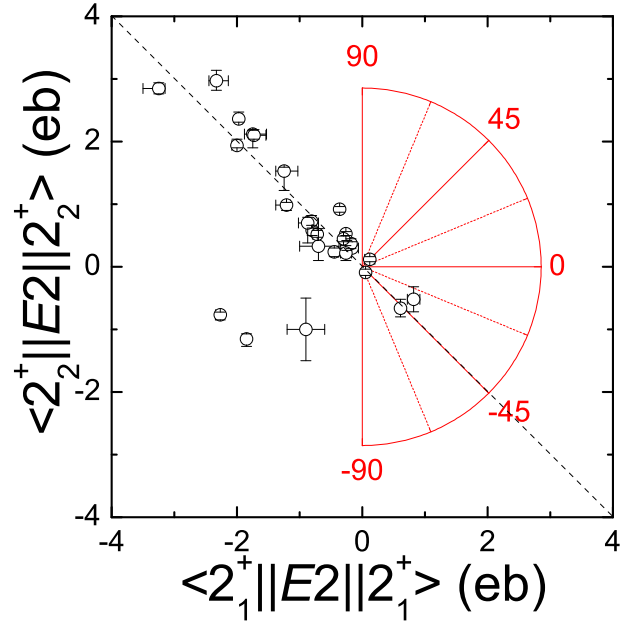}
\caption{(Color online) $\langle 2^+_1||E2||2^+_1\rangle$ and $\langle 2^+_2||E2||2^+_2\rangle$ matrix elements from Table I of ref. \cite{allmond} (i.e., [$Q(2^+_1)$, $Q(2^+_2)$] plots scaled by $\sqrt{16\pi/5}\langle 2220|22\rangle$). The $\theta$ parameterization defined in Eq. (\ref{theta}) is illustrated in the red sector. The $Q(2^+_2)= -Q(2^+_1)$ correlation is obvious along the $\theta= -45^{\circ}$ direction (the black dash diagonal line).}\label{q_exp}
\end{figure}

This work further studies the robust correlation between $Q$ values of the first two $2^+_1$ states, inspired by a recent experimental survey \cite{allmond}. As shown in Fig. \ref{q_exp}, this survey demonstrated a global $Q(2^+_2)= -Q(2^+_1)$ correlation across a wide range of masses, deformations, and $2^+_1$ energies. We will make use of random-interaction ensembles to provide an interacting-particle vision to this correlation, and search for other underlying $Q$ correlations. The statistic analysis based on the Elliott SU(3) model \cite{elliott-su3} and the mean-field Hartree-Bose theory \cite{ibm} is applied.

\section{calculation framework}\label{cal}
In our random-interaction calculations, the single-particle-energy degree of freedom is switched off to avoid the interference from the shell-structure detail. The two-body interaction matrix element, on the other hand, is denoted by $V_{j_1j_2j_3j_4}^J$ as usual, where $j_1$, $j_2$, $j_3$ and $j_4$ represent the angular momenta of single-particle orbits (half integer for fermions and integer for bosons), and the superscript $J$ labels the total angular momentum of the two-body configurations involved the interaction element. In our calculations, $V_{j_1j_2j_3j_4}^J$ is randomized independently and Gaussianly with $(\mu=0,~\sigma^2=1+\delta_{j_1j_2,j_3j_4})$, which insures the invariance of our random two-body interactions under arbitrary orthogonal transformations \cite{wigner}. All the possibilities of random interactions and their outputs via microscope-calculations construct the two-body random ensemble (TBRE) \cite{tbre-1,tbre-2,tbre-3}. Obviously, in the TBRE, diagonal interaction elements potentially have larger magnitudes.

For the shell-model TBRE in this work, four model spaces with either four or six valence protons in either $sd$ or $pf$ shell are considered, correspondingly to four nuclei: $^{24}$Si, $^{26}$S, $^{44}$Cr and $^{46}$Fe. For the IBM1 TBRE, $sd$-boson spaces are constructed for nuclei with valence boson numbers $N_b=$12, 13, 14 and 15, where $s$ and $d$ represents $I=0\hbar$ and $I=2\hbar$ bosons, respectively. It is noteworthy that a single calculation with random interactions does not match, and does not intend to match, to a realistic nucleus. It only presents a pseudo nucleus in the computational laboratory. Thus, in this article, model spaces described above are named as corresponding pseudo nuclei for convenience. For example, the model space with four protons in the $sd$ shell corresponds to pseudo $^{24}$Si. Statistic properties of many random-interaction calculations for pseudo nuclei can be related to the robustness of dynamic features in realistic nuclear systems. To insure the statistic validity of our conclusions, 1 000 000 sets of random interactions are generated for each pseudo nucleus, and inputted into the shell-model or IBM1 calculations. If one calculation produces a $I=0$ ground state, $Q$ matrix elements of $2^+_1$ and $2^+_2$ states will be calculated and recorded for the following statistic analysis. 

\section{$Q$ correlations in Shell Model}\label{sm}

In the Shell Model, the $Q$ matrix element of one $2^+$ state, $|2^+\rangle$, is defined conventionally as
\begin{equation}
\begin{aligned}
&Q(2^+)=\langle 2^+||\hat{Q}||2^+ \rangle,\\
&\hat{Q}=\langle j||r^2Y^2||j^{\prime}\rangle(a^{\dagger}_j\times \tilde{a}_{j^{\prime}})^{(2)},
\end{aligned}
\end{equation}
where $a^{\dagger}_j$ and $\tilde{a}_{j^{\prime}}$ are single-particle creation and time-reversal operators at orbits $j$ and $j^{\prime}$, respectively. A proportional $Q$ correlation between the first two $2^+_1$ states is normally characterized by the ratio of $Q(2^+_2)/Q(2^+_1)$. Geometrically, such correlation also corresponds to a straight line with the polar angle, 
\begin{equation}\label{theta}
\theta= \arctan\left\{\frac{Q(2^+_2)}{Q(2^+_1)}\right\},
\end{equation}
across the origin in the [$Q(2^+_1)$, $Q(2^+_2)$] plane. For example, the experimental $Q(2^+_1)=-Q(2^+_2)$ correlation suggested by Ref. \cite{allmond} can be illustrated by a diagonal $\theta=\arctan(-1)=-45^{\circ}$ line as expected in Fig. \ref{q_exp}. We also visualize the polar-angle scheme of the [$Q(2^+_1)$, $Q(2^+_2)$] plane in Fig. \ref{q_exp}.  

In this work, we prefer the statistic analysis based on the polar angle $\theta$ over the $Q(2^+_2)/Q(2^+_1)$ ratio because of two reasons. Firstly, the distribution of the $Q(2^+_2)/Q(2^+_1)$ ratio spreads widely, so that the statistic detail about $Q(2^+_2)=-Q(2^+_1)$ correlation may be concealed. In particular, there robustly exists 8\% probability of $|Q(2^+_2)/Q(2^+_1)|>10$ due to the predominance of weak quadruple collectivity, i.e. small $|Q(2^+_1)|$, in the shell-model TBRE \cite{horoi-be2}. However, we intend as comprehensively as possible to present the statistic detail about the experimental $Q(2^+_2)=-Q(2^+_1)=-1$ correlation. The wide statistic range of the $Q(2^+_2)/Q(2^+_1)$ ratio may fails this intention. By converting the $Q(2^+_2)/Q(2^+_1)$ ratio to the $\theta$ value, the statistic range is limited between $(-90^{\circ},90^{\circ})$, and a clearer vision around the $Q(2^+_2)=-Q(2^+_1)$ correlation can be obtained around $\theta=-45^{\circ}$. Secondly, the $\theta$ parameterization intuitively provides a reasonable geometric standard of symmetric sampling. Taking the $Q(2^+_2)=-Q(2^+_1)$ correlation for example, there is actually no (pseudo) nucleus following exact $Q(2^+_2)=-Q(2^+_1)$ relation in experiments or our TBRE, and yet we can take $\theta\in(-50^{\circ},-40^{\circ})$ as the sampling range to represent this correlation. One sees this sampling range indeed covers a symmetric area related to the $Q(2^+_2)=-Q(2^+_1)$ correlation in the [$Q(2^+_1)$, $Q(2^+_2)$] plane. With the $Q(2^+_2)/Q(2^+_1)$ statistic, the determination of symmetric sampling range for one specifically $Q$ correlation can be controversial or simply another representation of the $\theta$ parameterization. Therefore, all the statistic, analyses, and discussions in this work are based on the $\theta$ value.

\begin{figure}
\includegraphics[angle=0,width=0.45\textwidth]{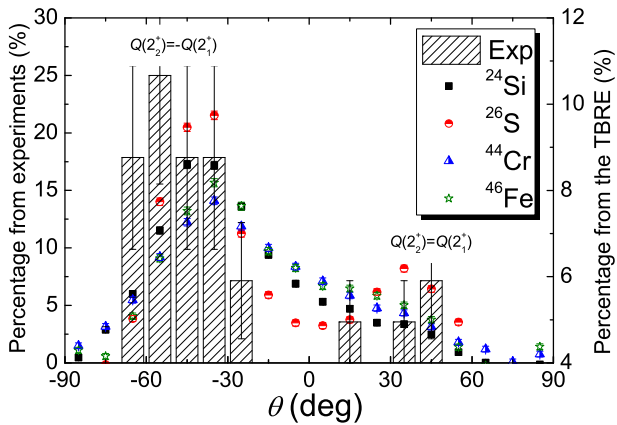}
\caption{(Color online) $\theta$ distributions from the experimental survey (Exp) \cite{allmond} and the shell-model TBRE. $\theta= \pm 45^{\circ}$ peaks are highlighted correspondingly to $Q(2^+_2)= \pm Q(2^+_1)$ correlations, respectively. Error bars correspond to statistic error.}\label{q_sm}
\end{figure}

In Fig. \ref{q_sm}, we present $\theta$ distributions of four pseudo nuclei in the shell-model TBRE compared with the experimental distribution from Ref. \cite{allmond}. The experimental $Q(2^+_1)=-Q(2^+_2)$ correlation is represented by the main peak around $\theta=-45^{\circ}$, which is also reproduced by the TBRE. Furthermore, several weak peaks around $\theta= 45^{\circ}$ are also observed in both experimental data and random-interaction systems, corresponding to the $Q(2^+_2)= Q(2^+_1)$ correlation.

As proposed by Ref. \cite{allmond}, nuclear rotor models can give the $\theta=-45^{\circ}$ correlation, even although such correlation experimentally occurs in both rotational or non-rotational nuclei. Therefore, we will further examine whether $\theta=\pm45^{\circ}$ correlations is the symbol of the underlying rotational collectivity in TBRE. Firstly, we verify whether $\theta=\pm 45^{\circ}$ correlations accompany rotational spectra in the TBRE. Secondly, we search statistic signature of the random-interaction elements that provides the $\theta=\pm 45^{\circ}$ correlations, and trace such signature back to the microscopical Hamiltonian of nuclear rotor model, namely the Elliott SU(3) Hamiltonian.

\begin{figure}
\includegraphics[angle=0,width=0.45\textwidth]{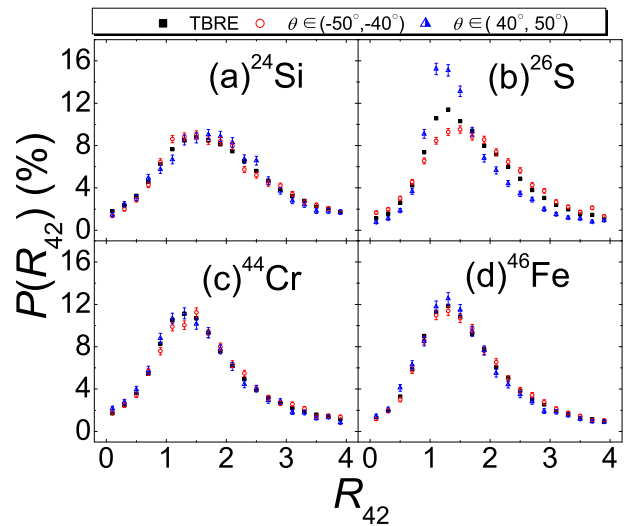}
\caption{(Color online) $R_{42}$ distributions around $\theta=\pm 45^{\circ}$ correlations (red circles and blue triangles, respectively) compared with those in the whole shell-model TBRE (black squares). Error bars correspond to statistic error.}\label{r_sm}
\end{figure}

Following previous random-interaction studies \cite{bijker-prl,bijker-prc,zhao-be2,bijker-mf}, potential rotational spectra with $\theta=\pm 45^{\circ}$ correlations can be characterized by the energy ratio $R_{42}=E_{4^+_1}/E_{2^+_1}\simeq10/3$, where $E_{2^+_1}$ and $E_{4^+_1}$ correspond to the excitation energy of yrast $2^+$ and $4^+$ states, respectively. Thus, we plot the $R_{42}$ distributions with $\theta\in (-50^{\circ},-40^{\circ})$ and $\theta\in (40^{\circ},50^{\circ})$, respectively, in Fig. \ref{r_sm}, and compare them with that in the whole TBRE. Except for $^{26}$S, $R$ distributions in both $\theta\in (-50^{\circ},-40^{\circ})$ and $\theta\in (40^{\circ},50^{\circ})$ regions are identical to those in the whole TBRE within statistic error. For $^{26}$S, the $R_{42}$ distribution has an observable enhancement at $R_{42}=1$ with $\theta\in (40^{\circ},50^{\circ})$. Namely, the $\theta=45^{\circ}$ correlation seems to partially originate from the seniority-like level scheme in $^{26}$S space. This observation explains why the $\theta=45^{\circ}$ peak for $^{26}$S is stronger shown in Fig. \ref{q_sm}, given the dominance of pairing-like behaviors in the TBRE \cite{johnson-prl,johnson-prc,rand-sen}. Nevertheless, there is no special favor on rotational spectra from $\theta=\pm 45^{\circ}$ correlations in the shell-model TBRE, consistently with the survey on the realistic nuclear system \cite{allmond}.

\begin{table}
\caption{$|\overline{V^J_{j_1j_2j_3j_4}}|$ values around $\theta \pm 45^{\circ}$ correlations and $|\langle j_1j_2|\hat{C}_{\rm SU(3)}|j_3j_4\rangle^J|$ elements [``SU(3)" column] in the $sd$ shell. The ``index" column presents the integer, $2j_1\times 10000+2j_2\times 1000+2j_3\times 100+2j_4\times 10+J$, to identify two-body interaction elements. All the data is organized in an increasing order of the index column.}\label{int-sd}
\begin{tabular}{cccccccccccccccccccccc}
\hline\hline
\multirow{2}{*}{Order} & $~~~$ & \multirow{2}{*}{Index} & $~$ & \multicolumn{2}{c}{$|\overline{V^J_{j_1j_2j_3j_4}}|$} & $~$ & \multirow{2}{*}{SU(3)} \\
\cline{5-6} 
& & & & $\theta\in(-50^{\circ},-40^{\circ})$ & $\theta\in(40^{\circ},50^{\circ})$ & & \\
\hline 
1 & & 11110 & & 0.031 & 0.028 & & 20.0 \\
2 & & 11330 & & 0.009 & 0.006 & & 5.7 \\
3 & & 11550 & & 0.017 & 0.006 & & 6.9 \\
4 & & 13131 & & 0.047 & 0.007 & & 7.0 \\
5 & & 13132 & & 0.022 & 0.053 & & 10.2 \\
6 & & 13152 & & 0.007 & 0.028 & & 3.9 \\
7 & & 13332 & & 0.035 & 0.025 & & 2.5 \\
8 & & 13351 & & 0.006 & 0.007 & & 0.0 \\
9 & & 13352 & & 0.002 & 0.043 & & 2.3 \\
10 & & 13552 & & 0.006 & 0.024 & & 3.3 \\
11 & & 15152 & & 0.045 & 0.087 & & 11.8 \\
12 & & 15153 & & 0.095 & 0.172 & & 7.0 \\
13 & & 15332 & & 0.024 & 0.005 & & 3.1 \\
14 & & 15352 & & 0.016 & 0.014 & & 2.8 \\
15 & & 15353 & & 0.004 & 0.060 & & 0.0 \\
16 & & 15552 & & 0.064 & 0.002 & & 4.0 \\
17 & & 33330 & & 0.118 & 0.039 & & 15.0 \\
18 & & 33332 & & 0.136 & 0.107 & & 0.7 \\
19 & & 33352 & & 0.025 & 0.021 & & 4.1 \\
20 & & 33550 & & 0.014 & 0.014 & & 2.4 \\
21 & & 33552 & & 0.016 & 0.018 & & 0.4 \\
22 & & 35351 & & 0.018 & 0.076 & & 13.0 \\
23 & & 35352 & & 0.015 & 0.104 & & 7.2 \\
24 & & 35353 & & 0.180 & 0.009 & & 2.0 \\
25 & & 35354 & & 0.028 & 0.167 & & 2.0 \\
26 & & 35552 & & 0.015 & 0.028 & & 4.8 \\
27 & & 35554 & & 0.046 & 0.021 & & 0.0 \\
28 & & 55550 & & 0.151 & 0.086 & & 16.0 \\
29 & & 55552 & & 0.046 & 0.069 & & 8.1 \\
30 & & 55554 & & 0.160 & 0.037 & & 2.0 \\
\hline\hline
\end{tabular}
\end{table}

\begin{table*}
\caption{The same as Table \ref{int-sd} except for the $pf$ shell.}\label{int-pf}
\begin{tabular}{cccccccccccccccccccccc}
\hline\hline
\multirow{2}{*}{Order} & $~$ & \multirow{2}{*}{Index} & $~$ & \multicolumn{2}{c}{$|\overline{V^J_{j_1j_2j_3j_4}}|$} & $~$ & \multirow{2}{*}{SU(3)} & $~~~~~~~~~$ & \multirow{2}{*}{Order} & $~$ & \multirow{2}{*}{Index} & $~$ & \multicolumn{2}{c}{$|\overline{V^J_{j_1j_2j_3j_4}}|$} & $~$ & \multirow{2}{*}{SU(3)} \\
\cline{5-6}\cline{14-15} 
& & & & $\theta\in(-50^{\circ},-40^{\circ})$ & $\theta\in(40^{\circ},50^{\circ})$ & & & & & & & & $\theta\in(-50^{\circ},-40^{\circ})$ & $\theta\in(40^{\circ},50^{\circ})$ & & \\
\hline 
1 & & 11110 & & 0.030 & 0.011 & & 32.0 & & 48 & & 35352 & & 0.059 & 0.110 & & 18.9 \\
2 & & 11330 & & 0.011 & 0.008 & & 12.2 & & 49 & & 35353 & & 0.044 & 0.030 & & 9.1 \\
3 & & 11550 & & 0.010 & 0.006 & & 9.7 & & 50 & & 35354 & & 0.044 & 0.011 & & 12.3 \\
4 & & 11770 & & 0.006 & 0.007 & & 0.0 & & 51 & & 35372 & & 0.002 & 0.001 & & 2.3 \\
5 & & 13131 & & 0.018 & 0.033 & & 23.4 & & 52 & & 35373 & & 0.004 & 0.005 & & 3.8 \\
6 & & 13132 & & 0.039 & 0.004 & & 27.9 & & 53 & & 35374 & & 0.006 & 0.007 & & 3.8 \\
7 & & 13152 & & 0.000 & 0.000 & & 6.6 & & 54 & & 35552 & & 0.003 & 0.011 & & 3.7 \\
8 & & 13332 & & 0.009 & 0.007 & & 3.2 & & 55 & & 35554 & & 0.010 & 0.000 & & 3.3 \\
9 & & 13351 & & 0.000 & 0.000 & & 0.0 & & 56 & & 35571 & & 0.006 & 0.000 & & 0.0 \\
10 & & 13352 & & 0.002 & 0.000 & & 3.5 & & 57 & & 35572 & & 0.002 & 0.002 & & 4.1 \\
11 & & 13372 & & 0.007 & 0.004 & & 8.6 & & 58 & & 35573 & & 0.010 & 0.004 & & 6.2 \\
12 & & 13552 & & 0.003 & 0.005 & & 4.1 & & 59 & & 35574 & & 0.005 & 0.002 & & 5.1 \\
13 & & 13571 & & 0.008 & 0.000 & & 9.0 & & 60 & & 35772 & & 0.004 & 0.001 & & 3.3 \\
14 & & 13572 & & 0.007 & 0.005 & & 5.4 & & 61 & & 35774 & & 0.004 & 0.005 & & 2.6 \\
15 & & 13772 & & 0.004 & 0.002 & & 0.0 & & 62 & & 37372 & & 0.035 & 0.032 & & 26.8 \\
16 & & 15152 & & 0.006 & 0.006 & & 20.1 & & 63 & & 37373 & & 0.033 & 0.041 & & 15.3 \\
17 & & 15153 & & 0.077 & 0.079 & & 6.5 & & 64 & & 37374 & & 0.067 & 0.058 & & 12.9 \\
18 & & 15173 & & 0.002 & 0.003 & & 1.0 & & 65 & & 37375 & & 0.060 & 0.042 & & 6.0 \\
19 & & 15332 & & 0.001 & 0.000 & & 4.7 & & 66 & & 37552 & & 0.001 & 0.004 & & 0.9 \\
20 & & 15352 & & 0.005 & 0.005 & & 7.9 & & 67 & & 37554 & & 0.000 & 0.008 & & 1.0 \\
21 & & 15353 & & 0.001 & 0.009 & & 1.3 & & 68 & & 37572 & & 0.004 & 0.002 & & 2.4 \\
22 & & 15372 & & 0.000 & 0.001 & & 2.8 & & 69 & & 37573 & & 0.001 & 0.002 & & 3.3 \\
23 & & 15373 & & 0.000 & 0.009 & & 1.9 & & 70 & & 37574 & & 0.007 & 0.000 & & 1.5 \\
24 & & 15552 & & 0.025 & 0.000 & & 6.1 & & 71 & & 37575 & & 0.005 & 0.004 & & 0.0 \\
25 & & 15572 & & 0.003 & 0.000 & & 2.4 & & 72 & & 37772 & & 0.010 & 0.003 & & 3.4 \\
26 & & 15573 & & 0.008 & 0.000 & & 2.2 & & 73 & & 37774 & & 0.026 & 0.000 & & 8.6 \\
27 & & 15772 & & 0.001 & 0.000 & & 0.0 & & 74 & & 55550 & & 0.071 & 0.073 & & 26.4 \\
28 & & 17173 & & 0.069 & 0.097 & & 18.9 & & 75 & & 55552 & & 0.041 & 0.040 & & 12.2 \\
29 & & 17174 & & 0.045 & 0.074 & & 12.0 & & 76 & & 55554 & & 0.042 & 0.019 & & 7.1 \\
30 & & 17353 & & 0.000 & 0.002 & & 1.1 & & 77 & & 55572 & & 0.007 & 0.004 & & 5.8 \\
31 & & 17354 & & 0.011 & 0.005 & & 5.9 & & 78 & & 55574 & & 0.003 & 0.000 & & 3.6 \\
32 & & 17373 & & 0.014 & 0.003 & & 8.9 & & 79 & & 55770 & & 0.016 & 0.013 & & 2.1 \\
33 & & 17374 & & 0.015 & 0.002 & & 2.0 & & 80 & & 55772 & & 0.009 & 0.005 & & 1.2 \\
34 & & 17554 & & 0.002 & 0.002 & & 3.4 & & 81 & & 55774 & & 0.002 & 0.010 & & 0.2 \\
35 & & 17573 & & 0.010 & 0.002 & & 4.8 & & 82 & & 57571 & & 0.006 & 0.026 & & 24.6 \\
36 & & 17574 & & 0.013 & 0.000 & & 6.3 & & 83 & & 57572 & & 0.067 & 0.080 & & 18.0 \\
37 & & 17774 & & 0.000 & 0.002 & & 0.0 & & 84 & & 57573 & & 0.053 & 0.015 & & 7.2 \\
38 & & 33330 & & 0.070 & 0.045 & & 40.6 & & 85 & & 57574 & & 0.028 & 0.004 & & 0.2 \\
39 & & 33332 & & 0.101 & 0.126 & & 25.6 & & 86 & & 57575 & & 0.095 & 0.045 & & 9.0 \\
40 & & 33352 & & 0.005 & 0.005 & & 2.5 & & 87 & & 57576 & & 0.020 & 0.036 & & 9.0 \\
41 & & 33372 & & 0.007 & 0.001 & & 6.1 & & 88 & & 57772 & & 0.010 & 0.006 & & 5.4 \\
42 & & 33550 & & 0.011 & 0.021 & & 2.0 & & 89 & & 57774 & & 0.010 & 0.000 & & 5.1 \\
43 & & 33552 & & 0.010 & 0.010 & & 0.2 & & 90 & & 57776 & & 0.013 & 0.002 & & 0.0 \\
44 & & 33572 & & 0.004 & 0.001 & & 4.4 & & 91 & & 77770 & & 0.082 & 0.071 & & 27.0 \\
45 & & 33770 & & 0.009 & 0.019 & & 10.2 & & 92 & & 77772 & & 0.025 & 0.011 & & 18.6 \\
46 & & 33772 & & 0.005 & 0.004 & & 6.7 & & 93 & & 77774 & & 0.092 & 0.080 & & 3.1 \\
47 & & 35351 & & 0.004 & 0.027 & & 27.0 & & 94 & & 77776 & & 0.123 & 0.118 & & 9.0 \\
\hline\hline
\end{tabular}
\end{table*}

In Ref. \cite{horoi-a}, the interaction signature of prolate and oblate shapes is represented by the average values of interaction elements (denoted by $\overline{V^J_{j_1j_2j_3j_4}}$). In this work, we also adopt $\overline{V^J_{j_1j_2j_3j_4}}$ to probe the the interaction signature of $\theta=\pm 45^{\circ}$ correlations. In detail, we collect all the interaction elements within $\theta\in(-50^{\circ},-40^{\circ})$ and $\theta\in(40^{\circ},50^{\circ})$, normalize them by the factor of $\sum\limits_{Jj_1j_2j_3j_4} V^J_{j_1j_2j_3j_4}$, and then calculate all the $\overline{V^J_{j_1j_2j_3j_4}}$ values for both $\theta\in(-50^{\circ},-40^{\circ})$ and $\theta\in(40^{\circ},50^{\circ})$ regions, respectively. Because signs of interaction elements can be changed by different phase conventions, we only discuss magnitudes of $\overline{V^J_{j_1j_2j_3j_4}}$ (denoted by $|\overline{V^J_{j_1j_2j_3j_4}}|$) to avoid the potential ambiguity from phase conventions. To simplify the following discussion, each $|\overline{V^J_{j_1j_2j_3j_4}}|$ is labeled by the index, $2j_1\times 10000+2j_2\times 1000+2j_3\times 100+2j_4\times 10+J$. For example, the pairing force between $s_{1/2}$ or $p_{1/2}$ nucleons, $V^0_{\frac{1}{2}\frac{1}{2}\frac{1}{2}\frac{1}{2}}$, corresponds to index ``11110". We list $|\overline{V^J_{j_1j_2j_3j_4}}|$ values of both $\theta\in(-50^{\circ},-40^{\circ})$ and $\theta\in(40^{\circ},50^{\circ})$ region in an increasing order of their indices in Tables \ref{int-sd} and \ref{int-pf}.

\begin{figure*}
\includegraphics[angle=0,width=0.8\textwidth]{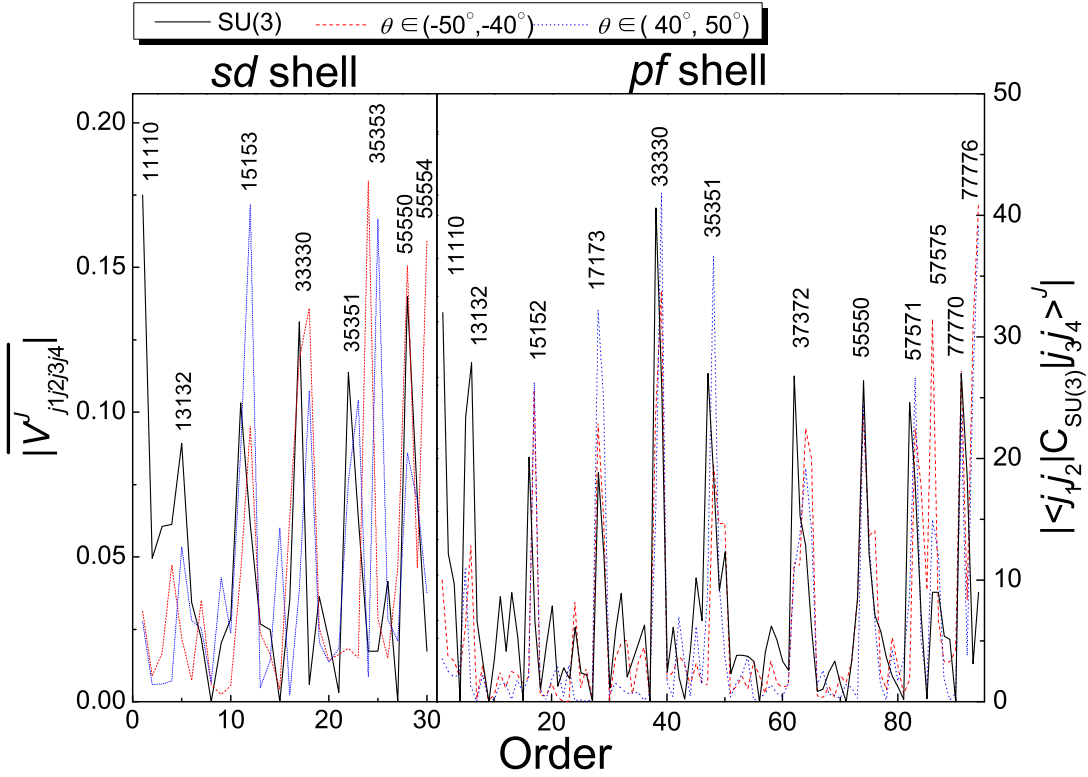}
\caption{(Color online) $|\overline{V^J_{j_1j_2j_3j_4}}|$ and $|\langle (j_1j_2)^J|\hat{C}_{\rm SU(3)}|(j_3j_4)^J\rangle|$ values (see text for definitions) against order numbers from Tables \ref{int-sd} and \ref{int-pf}. indices are highlighted for obvious peaks for $|\langle (j_1j_2)^J|\hat{C}_{\rm SU(3)}|(j_3j_4)^J\rangle|$ values.}\label{int_mean}
\end{figure*}

To comprehensively compare $|\overline{V^J_{j_1j_2j_3j_4}}|$ values between $\theta\in(-50^{\circ},-40^{\circ})$ and $\theta\in(40^{\circ},50^{\circ})$ regions, we plot them against their order numbers (see Table \ref{int-sd} and \ref{int-pf}) in Fig. \ref{int_mean}. Most of $\overline{V^J_{j_1j_2j_3j_4}}$ are close to zero following the ensemble distribution. However, there are several relatively large $|\overline{V^J_{j_1j_2j_3j_4}}|$ values, which presents obvious peaks in Fig. \ref{int_mean}. Peak positions for $\theta\in(-50^{\circ},-40^{\circ})$ are roughly consistent with those for $\theta\in(40^{\circ},50^{\circ})$, which hints that $\theta=\pm 45^{\circ}$ correlations may share the same interaction signature. 

The interaction signature of $\theta=\pm 45^{\circ}$ correlations can be related to the Elliott Hamiltonian. Such Hamiltonian is dominated by the SU(3) Casimir operator as defined by
\begin{equation}\label{h-su3}
\hat{C}_{\rm SU(3)}=\frac{1}{4}\hat{Q}\cdot \hat{Q}+\frac{3}{4}\hat{L}\cdot \hat{L},
\end{equation}
where $\hat{Q}$ and $\hat{L}$ are quadrupole-moment and orbital-angular-momentum operators. We calculate matrix elements of $\langle (j_1j_2)^J|\hat{C}_{\rm SU(3)}|(j_3j_4)^J\rangle$, and still focus on their magnitudes (denoted by $|\langle j_1j_2|\hat{C}_{\rm SU(3)}|j_3j_4\rangle^J|$), similarly to the treatment for $\overline{V^J_{j_1j_2j_3j_4}}$. $|\langle j_1j_2|\hat{C}_{\rm SU(3)}|j_3j_4\rangle^J|$ is also labeled by index, $2j_1\times 10000+2j_2\times 1000+2j_3\times 100+2j_4\times 10+J$, and thus comparable with $|\overline{V^J_{j_1j_2j_3j_4}}|$ as shown in Table \ref{int-sd}, \ref{int-pf} and Fig. \ref{int_mean}. In Fig. \ref{int_mean}, relatively large $|\langle j_1j_2|\hat{C}_{\rm SU(3)}|j_3j_4\rangle^J|$ also presents several obvious peaks, which have similar pattern to $|\overline{V^J_{j_1j_2j_3j_4}}|$ peaks for both $\theta\in(-50^{\circ},-40^{\circ})$ and $\theta\in(40^{\circ},50^{\circ})$ regions. This observation implies the relation between the SU(3) symmetry and $\theta=\pm 45^{\circ}$ correlations. 

We also highlight indices for $|\langle j_1j_2|\hat{C}_{\rm SU(3)}|j_3j_4\rangle^J|$ peaks in Fig. \ref{int_mean}, according to which, the SU(3) Casimir operator always has large magnitudes for diagonal matrix elements with $j_1j_2=j_3j_4$. On the other hand, large $|\overline{V^J_{j_1j_2j_3j_4}}|$ for $\theta=\pm 45^{\circ}$ correlations also occurs for diagonal $j_1j_2=j_3j_4$ in Table \ref{int-sd} and \ref{int-pf}. As described in Sec. \ref{cal}, larger magnitudes of diagonal elements is required by the invariance of TBRE under orthogonal transformation of two-body configuration. Therefore, the shell-model TBRE intrinsically maintains part of the SU(3) properties to restore the $\theta=\pm 45^{\circ}$ correlations, even though it spectrally presents no trace of the SU(3) symmetry as illustrated in Fig. \ref{r_sm}.

After clarifying the relation between $\theta=\pm 45^{\circ}$ correlations and the SU(3) symmetry, we microscopically describe how these two $Q$ correlations emerge in a major shell, i.e. $sd$ or $pf$ shell here. In the Elliott model, any $2^+$ state within a major shell is labeled by the SU(3) representation $(\lambda,~\mu)$, the quantum number of the intrinsic state ($K$), and orbital angular momentum $L=2$ \cite{elliott-su3}. The $2^+$ state is normally near the bottom of a $K$ band, and thus its $Q$ value can be approximately given by \cite{elliott-q}
\begin{equation}
Q(2^+)=\frac{2\lambda}{7}(K^2-2).
\end{equation}
The $K$ number is limited to 0, 1 and 2. Thus, the $Q(2^+_1)=-Q(2^+_2)$, i.e. $\theta=-45^{\circ}$, correlation is produced by two $2^+$ states with the same $\lambda$ number and $K=0,~2$ respectively, which agrees with the rotor-model conjecture \cite{allmond}. On the other hand, the $Q(2^+_1)=Q(2^+_2)$, i.e. $\theta=45^{\circ}$, correlation is from two $2^+$ states with the same $\lambda$ and $K$ values. 

According to above SU(3) description, one can expect two $2^+$ states with the $\theta=-45^{\circ}$ correlation from the same $(\lambda,\mu)$ representation. On the contrary, a single $(\lambda,\mu)$ representation can not produced two $2^+$ states with the same $K$ number, so that the $\theta=45^{\circ}$ correlation always requires the cooperation of two different $(\lambda,\mu)$ representations. Empirically, the former case has a relatively larger probability to emerge in the low-lying region, which explains why the $\theta=-45^{\circ}$ peak intensity is always larger than the $\theta=45^{\circ}$ one in Fig. \ref{q_sm}.

Independently of the rotor interpretation, the anharmonic vibration (AHV) with quadrupole degrees of freedom \cite{ahv} can also provide the $\theta=-45^{\circ}$ correlation. In the AHV interpretation, the first two $2^+$ states are constructed with a significant mixing of one- and two-phonon configurations as
\begin{equation}\label{2+}
\begin{aligned}
|2^+_1\rangle&=a_1 |b^{\dagger}\rangle+a_2| (b^{\dagger})^2 \rangle,\\
|2^+_2\rangle&=-a_2 |b^{\dagger}\rangle+a_1 |(b^{\dagger})^2 \rangle,
\end{aligned}
\end{equation}
where $b^{\dagger}$ is the creation operator of a phonon; $a_1$ and $a_2$ are amplitudes of phonon configurations. In this phonon space, the quadrupole operator $\hat{Q}$ is a polynomial of operator $b^{\dagger}+\tilde{b}$ \cite{ahv-q}, where $\tilde{b}$ is the phonon time-reversal operator. The first order of such polynomial dominates the $Q$ matrix element. However, it also vanishes with respect to configuration with definite numbers of phonons. In particular,
\begin{equation}
\begin{aligned}
&\langle \tilde{b}||\hat{Q}||b^{\dagger}\rangle \propto \langle \tilde{b}||b^{\dagger}+\tilde{b}||b^{\dagger} \rangle=0,\\
&\langle (\tilde{b})^2||\hat{Q}||(b^{\dagger})^2\rangle\propto \langle (\tilde{b})^2||b^{\dagger}+\tilde{b}||(b^{\dagger})^2 \rangle=0,
\end{aligned}
\end{equation}
Thus, 
\begin{equation}
\begin{aligned}
&\langle2^+_1||\hat{Q}||2^+_1\rangle=2a_1a_2\langle \tilde{b}||\hat{Q}||(b^{\dagger})^2\rangle,\\
&\langle2^+_2||\hat{Q}||2^+_2\rangle=-2a_1a_2\langle \tilde{b}||\hat{Q}||(b^{\dagger})^2\rangle,
\end{aligned}
\end{equation}
and the $\theta=-45^{\circ}$ relation is obtained.

\begin{figure}
\includegraphics[angle=0,width=0.48\textwidth]{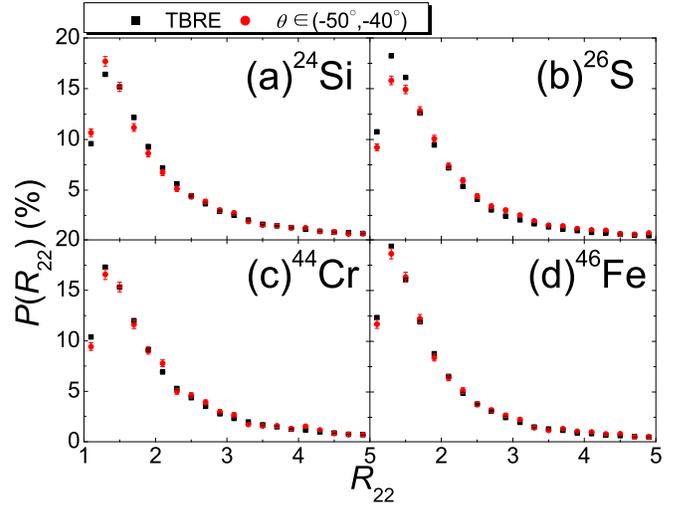}
\caption{(Color online) $R_{22}$ distributions around $\theta=- 45^{\circ}$ (red circles) compared with those in the whole shell-model TBRE (black squares). Error bars correspond to statistic error.}\label{r22_sm}
\end{figure}

We can spectrally examine this AHV interpretation for the $\theta=-45^{\circ}$ correlation in the shell-model TBRE. Because AHV $2^+$ states correspond to the mixing of one- and two-phonon configurations as defined in Eq. (\ref{2+}), the excitation energy of the first $2^+$state, $E(2^+_1)$, is smaller than the one-phonon excitation energy, $\hbar\omega$; while $E(2^+_2)$ is larger than $2\hbar\omega$, according to the perturbation theory. Thus, the energy ratio of $R_{22}=E(2^+_2)/E(2^+_1)$ of the AHV is always larger than 2. In other words, if the AHV contributes to the $\theta=-45^{\circ}$ correlation in the TBRE, the distribution of $R_{22}$ with $\theta\in(-50^{\circ},-40^{\circ})$ should have an obvious enhancement for $R_{22}>2$. In Fig. \ref{r22_sm}, we compare $R_{22}$ distributions in the $\theta\in(-50^{\circ},-40^{\circ})$ range and those in the whole shell-model TBRE. There is no obvious difference between these $R_{22}$ distributions. Thus, we don't see the spectral sign of the AHV contribution to the $\theta=-45^{\circ}$ correlation.

\section{$Q$ correlations in IBM1}
In IBM1, the $Q$ operator is a linear combination of two independent rank-two operators as:
\begin{equation}\label{q1q2_ibm}
Q=Q^1+\chi Q^2,
\end{equation}
where $Q^1=d^{\dagger}s+s\tilde{d}$, $Q^2=[d^{\dagger}\tilde{d}]^2$, and $\chi$ is a free parameter. Correspondingly, we need to define two independent $\theta$ coordinates as
\begin{equation}\label{theta_q1q2}
\begin{aligned}
\theta^1&= \arctan\left\{\frac{\langle 2^+_2||Q^1||2^+_2\rangle}{\langle 2^+_1||Q^1||2^+_1\rangle}\right\},\\
\theta^2&= \arctan\left\{\frac{\langle 2^+_2||Q^2||2^+_2\rangle}{\langle 2^+_1||Q^2||2^+_1\rangle}\right\}.
\end{aligned}
\end{equation}
A robust correlation with the polar angle $\theta$ should be insensitive to the $\chi$ value, which requires $\theta_1=\theta_2=\theta$. Obviously, such correlation corresponds to a peak at ($\theta$, $\theta$) point in the two-dimensional ($\theta^1$, $\theta^2$) distribution of the IBM1 TBRE.

\begin{figure*}
\includegraphics[angle=0,width=0.8\textwidth]{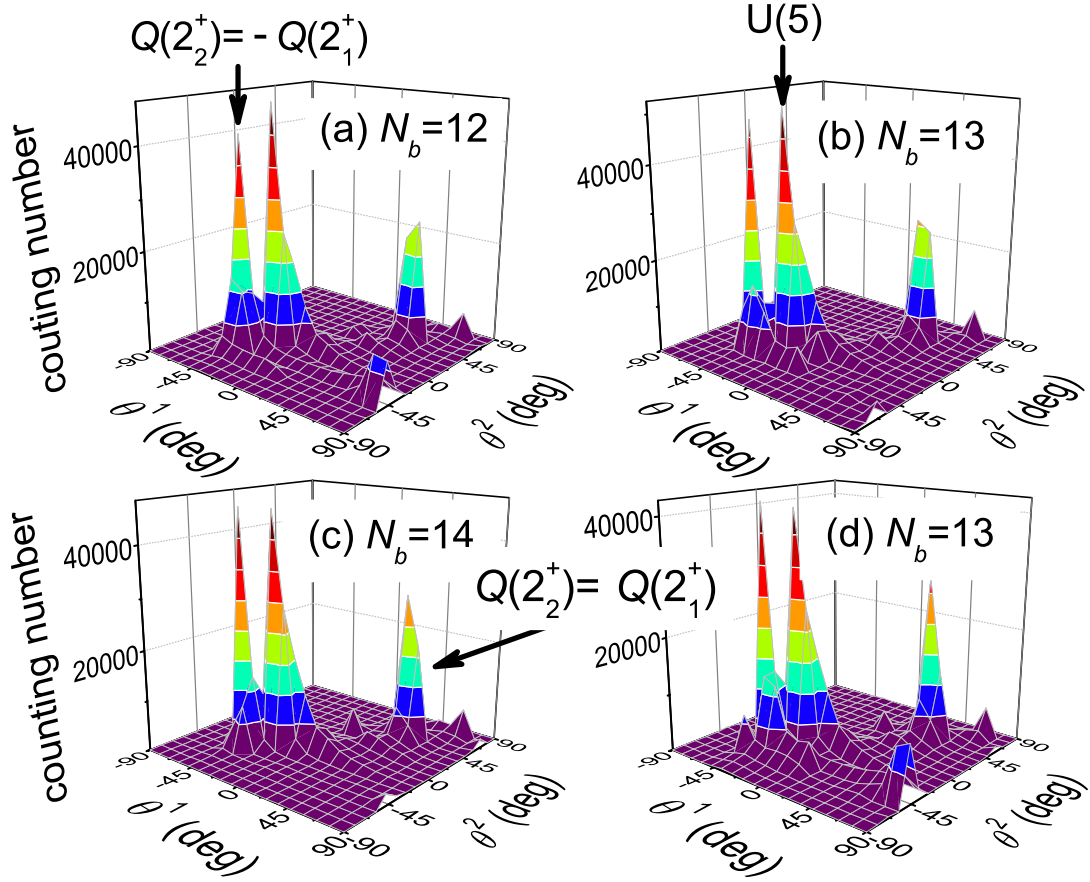}
\caption{(Color online) Two-dimensional ($\theta^1$, $\theta^2$) distributions of the IBM1 TBRE. Three sharp peaks are characterized with ``$Q(2^+_2)=\pm Q(2^+_1)$" and ``U(5)" corrlations.}\label{theta_ibm}
\end{figure*}

Fig. \ref{theta_ibm} represents ($\theta^1$, $\theta^2$) distributions of the IBM1 TBRE with $N_b=12$, 13, 14 and 15. These distributions follow similar pattern with three sharp peaks along the $\theta^1=\theta^2$ diagonal line, corresponding to three proportional $Q$ correlations. We fit ($\theta^1$, $\theta^2$) distributions to a two-dimensional function, $f(\theta^1,\theta^2)$, with three Gaussian peaks as
\begin{widetext}
\begin{equation}\label{gau}
f(\theta^1,\theta^2)=f_0+\sum\limits_{i=1}^3A_i \exp\left\{{-\frac{[(\theta^1-\theta^1_{c,i})\cos\omega_i+(\theta^2-\theta^2_{c,i})\sin\omega_i]^2}{2w^2_{\parallel , i}}}
{-\frac{[-(\theta^1-\theta^1_{c,i})\sin\omega_i+(\theta^2-\theta^2_{c,i})\cos\omega_i]^2}{2w^2_{\perp , i}}}\right\},
\end{equation}
\end{widetext}
where $f_0$ is the background; all the other fitting variables are parameters of Gaussian peaks. These three Gaussian peaks are labeled by indices $i=1$, 2 and 3. For the $i$th peak, $\omega_i$ defines its orientation in the $(\theta^1,\theta^2)$ plane, $(\theta^1_{c,i},\theta^2_{c,i})$ is the peak position, $A_i$ is the amplitude, and $(w_{\parallel , i}, w_{\perp , i})$ are widths along and perpendicularly to $\omega_i$ direction. Thus, the best-fit intensity of the $i$th peak can be calculated as $2\pi A_i w_{\parallel , i}w_{\perp , i}$.

\begin{table*}
\caption{Best-fit peak positions $(\theta^1_{c,i},\theta^2_{c,i})$ and intensities of three sharp peaks in Fig. \ref{theta_ibm} with the two-dimensional three-peak Gaussian function defined in Eq. (\ref{gau}).}\label{peak-fit}
\begin{tabular}{cccccccccccccccccccccc}
\hline\hline
\multirow{3}{*}{$N_b$} & $~~~$ & \multicolumn{3}{c}{$Q(2^+_2)=-Q(2^+_1)$} & $~~~$ & \multicolumn{3}{c}{$Q(2^+_2)=Q(2^+_1)$} & $~~~$ & \multicolumn{3}{c}{U(5)} \\
\cline{3-5}\cline{7-9}\cline{11-13} 
& & $\theta^1_{c,1}$ & $\theta^2_{c,1}$ & Intensity & & $\theta^1_{c,2}$ & $\theta^2_{c,2}$ & Intensity & & $\theta^1_{c,3}$ & $\theta^2_{c,3}$ & Intensity \\
& & (deg) & (deg) & ($\times 10^2$ counts) & & (deg) & (deg) & ($\times 10^2$ counts) & & (deg) & (deg) & ($\times 10^2$ counts) \\
\hline 
12 & & -42.05(1) & -36.27(2) & 361(6) & & 40.19(2) & 43.44(1) & 272(5) & & -21.27(4) & -22.17(1) & 565(9) \\
13 & & -42.29(1) & -36.92(2) & 476(6) & & 40.60(1) & 43.62(1) & 347(5) & & -21.12(3) & -22.21(1) & 742(9) \\
14 & & -42.52(1) & -37.45(1) & 468(6) & & 40.91(1) & 43.73(1) & 347(5) & & -21.20(2) & -22.23(1) & 733(8) \\
15 & & -42.66(1) & -37.94(1) & 393(7) & & 41.20(1) & 43.83(1) & 334(5) & & -21.32(2) & -22.26(1) & 683(8) \\
\hline\hline
\end{tabular}
\end{table*}

In Table \ref{peak-fit}, we list the best-fit peak positions and intensities for the three sharp peaks in Fig. \ref{theta_ibm}. The $i=1$ and $i=2$ peaks are very close to $(\pm 45^{\circ},\pm 45^{\circ})$, i.e. ``$Q(2^+_2)= \pm Q(2^+_1)$" correlations, as labeled in Fig. \ref{theta_ibm} and Table \ref{peak-fit}. The $i=3$ peak is located around $(-21^{\circ},-22^{\circ})$, and thus gives $Q(2^+_2)/Q(2^+_1)\simeq -3/7$, the typical IBM1 $Q$ ratio at the U(5) limit regardless of the boson number. Therefore, we believe the $i=3$ peak may correspond to the vibrational U(5) collectivity, and denote it as ``U(5)" in following analysis.

To identify or confirm the collective patterns corresponding to the three sharp peaks in Fig. \ref{theta_ibm}, we firstly investigate their $R_{42}$ distribution, i.e. the predominance of low-lying collective excitations, similarly to our $R_{42}$ analysis for the shell-model TBRE with Fig. \ref{r_sm}; secondly, we adopt the $sd$-boson mean-field theory to observe dominant nuclear sharps of these peaks.

\begin{figure}
\includegraphics[angle=0,width=0.48\textwidth]{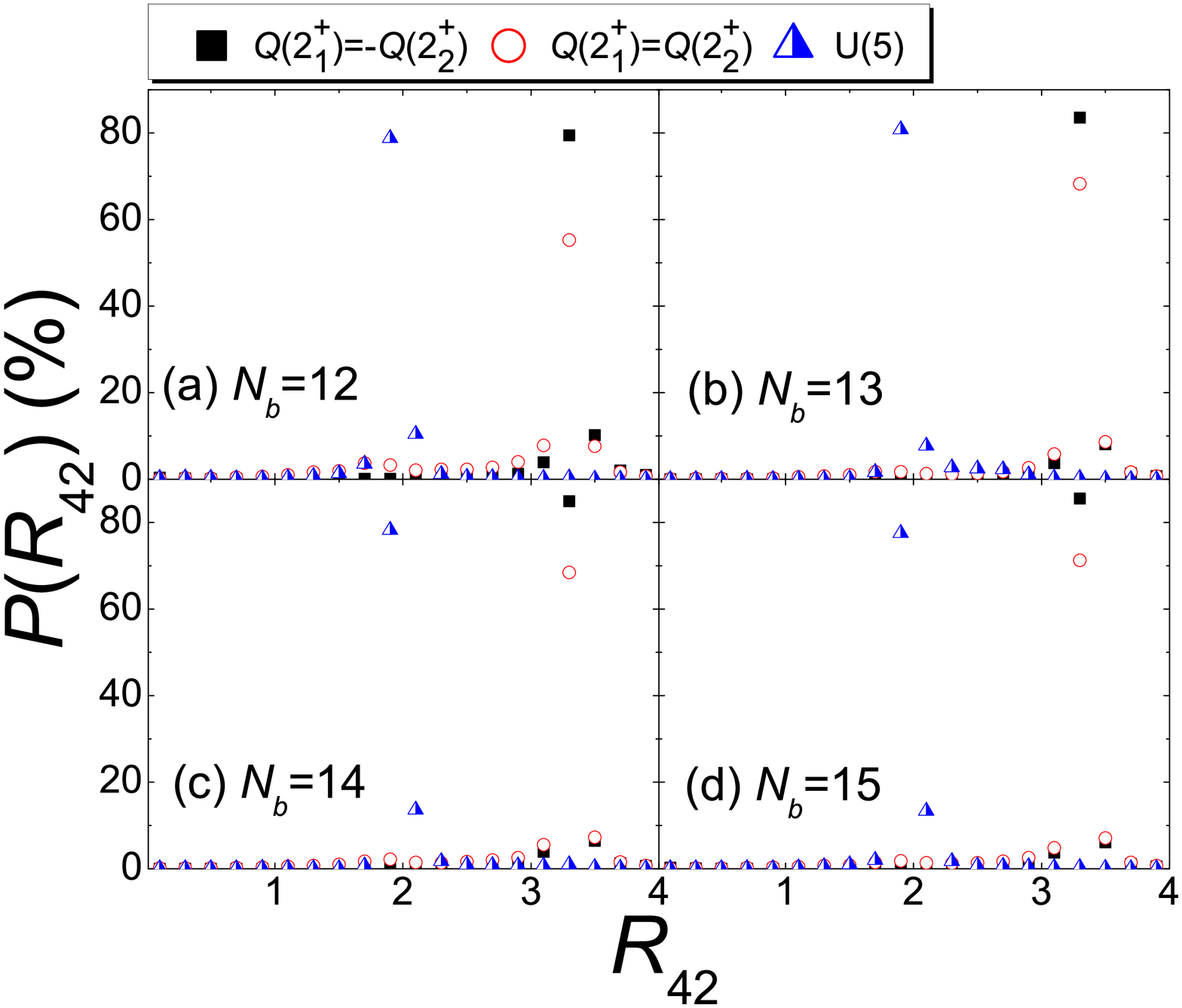}
\caption{(Color online) $R_{42}$ distributions around three peaks in Fig. \ref{theta_ibm}.}\label{r_ibm}
\end{figure}

For the analysis of $R_{42}$ distributions, we firstly collect all the random interactions, which produce ($\theta^1$, $\theta^2$) points within $3^{\circ}$ from peaks in Fig. \ref{theta_ibm}. Secondly, all $R_{42}$ values from these interactions are calculated. Thirdly, $R_{42}$ distributions of these peaks are calculated and presented in Fig. \ref{r_ibm}, respectively. $Q(2^+_2)= \pm Q(2^+_1)$ peaks always have large probabilities at rotational limit $R_{42}=3.3$, which agrees with the rotor-model description. On the other hand, $R_{42}$ distributions of U(5) peaks are dominated by $R_{42}=2$, corresponding to a typical U(5) vibrational spectrum, which supports our U(5) assignment for this peak.

Our analysis with the $sd$-boson mean-field theory starts with the $sd$-boson coherent state for the ground band as
\begin{equation}\label{coh}
|g\rangle=(s^{\dagger}+\tan \alpha_0 d^{\dagger}_0)^{N_b}|0\rangle.
\end{equation}
Similarly to Ref. \cite{bijker-mf}, the nuclear shape, i.e. the optimized $\alpha_0$ value, is determined by minimizing the Hamiltonian expectation value of this coherent state as
\begin{equation}\label{eg}
\begin{aligned}
E_g(\alpha)=&a_1\sin^4\alpha+a_2\sin^3\alpha\cos\alpha\\
&+a_3\sin^2\alpha\cos^2\alpha+a_0\cos^4\alpha,\\
\end{aligned}
\end{equation}
where $E_g(\alpha_0)$ reaches the minimum of this equation; $a_0$, $a_1$, $a_2$ and $a_3$ are linear combination of $sd$-boson two-body interaction matrix elements as formulated in Ref. \cite{chen-mf}. We calculate $\alpha_0$ values for all the interactions with spin-0$\hbar$ ground states in the TBRE, and perform frequency counting for calculated $\alpha_0$ values. Thus, the ensemble-normalized $\alpha$ distribution for the $i$th peaks is given by
\begin{equation}\label{palpha}
P(\alpha)=N(\alpha,\theta^1_{c,i},\theta^2_{c,i})/\mathcal{N}(\alpha),
\end{equation}
where $N(\alpha,\theta^1_{c,i},\theta^2_{c,i})$ is the counting number with $\alpha_0\in (\alpha-2.5^{\circ},\alpha+2.5^{\circ})$ and $\sqrt{(\theta^1-\theta^1_{c,i})^2+(\theta^2-\theta^2_{c,i})^2}<3^{\circ}$, and $\mathcal{N}(\alpha)$ is that with $\alpha_0\in (\alpha-2.5^{\circ},\alpha+2.5^{\circ})$ in the whole IBM1 TBRE.

\begin{figure}
\includegraphics[angle=0,width=0.48\textwidth]{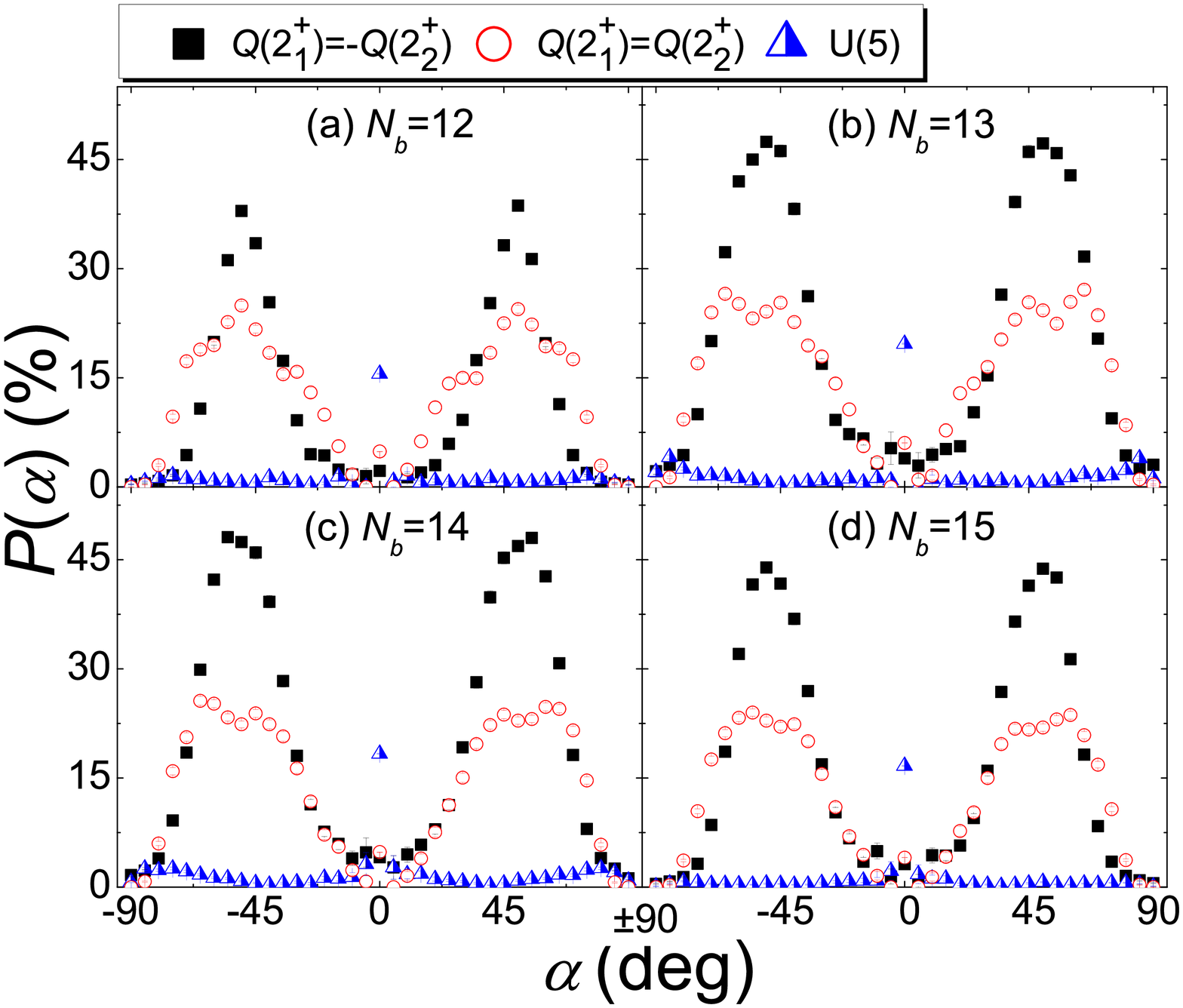}
\caption{(Color online) $P(\alpha)$ around three peaks in Fig. \ref{theta_ibm}, as defined in Eq. (\ref{palpha}).}\label{alpha}
\end{figure} 

Fig. \ref{alpha} presents calculated $P(\alpha)$s. The U(5) peak only has a significant probability at $\alpha=0$, corresponding to the $s$-boson condensation. Thus, $2^+$ states for the U(5) peak are constructed by replacing $s$ bosons with $d$ bosons in the $s$-boson condensation, which agrees with the quadrupole vibration described by the U(5) limit. This further confirms our U(5) characterization of this peak. On the other hand, the $Q(2^+_2)= \pm Q(2^+_1)$ peaks both have large probabilities for $0<|\alpha|<90^{\circ}$, corresponding to the axially symmetric rotor at the SU(3) limit. Considering that the $Q(2^+_2)= \pm Q(2^+_1)$ peaks also favor SU(3) rotational spectra with $R=3.3$ in Fig. \ref{r_ibm}, we conclude that both $Q(2^+_2)= \pm Q(2^+_1)$ correlations in IBM1 are strongly related to the SU(3) limit.

Conversely, we also derive $Q(2^+_2)= \pm Q(2^+_1)$ correlations from the SU(3) limit of the IBM1. At the SU(3) limit, the $2^+_1$ state is from the ground band with ($\lambda=2N_b$, $\mu=0$) and $K=0$; yet, the $2^+_2$ state belongs to the ($\lambda=2N_b-4$, $\mu=2$) representation, which generates $\beta$ and $\gamma$ bands with $K=0$ and 2, respectively \cite{ibm}. Thus, the $2^+_2$ is from either $\beta$ or $\gamma$ band, which leads to two phase-different $Q$ correlations 
\begin{equation}\label{q_nb}
\frac{Q^{\beta}(2^+)}{Q^{g}(2^+)}= \frac{4N_b-3}{4N_b+3};~\frac{Q^{\gamma}(2^+)}{Q^{g}(2^+)}= -\frac{4N_b-3}{4N_b+3}.
\end{equation}
For $N_b\rightarrow \infty$, $Q^{\beta}(2^+)= Q^{\rm g}(2^+)$ and $Q^{\gamma}(2^+)= -Q^{\rm g}(2^+)$ are achieved, corresponding to $Q(2^+_2)= Q(2^+_1)$ and $Q(2^+_2)= -Q(2^+_1)$ correlations, respectively.

In the shell-model TBRE, the $Q(2^+_2)= - Q(2^+_1)$ correlation has a larger probability than $Q(2^+_2)= Q(2^+_1)$ one (see Fig. \ref{q_sm}). Yet, in the IBM1 TBRE, these two correlations have roughly equal peak intensities, i.e. probabilities, as shown in Fig. \ref{theta_ibm} and Table \ref{peak-fit}. This is a major difference between behaviors of the $Q$ correlations in shell-model and IBM1 TBREs. This difference can be explained according to the $(\lambda, \mu)$ assignment of the SU(3) scheme. In the Shell Model, i.e. the Elliott model, the $Q(2^+_2)= - Q(2^+_1)$ correlation normally emerges with two $2^+$ states from a single $(\lambda, \mu)$ representation, which empirically provides a larger probability. However, in the $sd$-boson space, both $Q(2^+_2)= \pm Q(2^+_1)$ correlations require $2^+_1$ and $2^+_2$ states from two different $(\lambda, \mu)$ representations, and thus have similar probabilities.

As shown in Table \ref{peak-fit}, $|\theta^1_{c,i}|$ and $|\theta^2_{c,i}|$ of $Q(2^+_2)= \pm Q(2^+_1)$ correlations are systematically smaller than $45^{\circ}$. This observation can be explained with Eq. (\ref{q_nb}). For large but finite $N_b$, the magnitude of $Q(2^+_2)$ is always smaller that that of $Q(2^+_1)$, which drives the $|\theta^1_{c,i}|$ and $|\theta^2_{c,i}|$ value smaller than $45^{\circ}$. Therefore, we attribute the systematical derivation of $Q(2^+_2)\simeq \pm Q(2^+_1)$ peak positions from the exact SU(3) prediction to the finite-boson-number effect, as proposed by Ref. \cite{allmond} with consistent-$Q$ calculations.

\section{summary}
To summarize, we observe three proportional correlations between $Q$ values of the first two $I^{\pi}=2^+$ states in the TBRE. $Q(2^+_1)=\pm Q(2^+_2)$ correlations robustly and universally exists in both shell-model and $sd$ spaces, consistently with experiments. In the IBM1 TBRE, the $Q(2^+_2)=-\frac{3}{7}Q(2^+_1)$ correlation is also reported. By using the Elliot model and the $sd$-boson mean-field theory, we can microscopically assign $Q(2^+_1)=\pm Q(2^+_2)$ correlations to the rotational SU(3) symmetry, and the $Q(2^+_2)=-\frac{3}{7}Q(2^+_1)$ correlation to the quadrupole vibrational U(5) limit. Phenomenologically, the anharmonic vibration may also provide the $Q(2^+_1)=- Q(2^+_2)$ correlation, although its spectral behavior is not observed in the shell-model TBRE.

In particular, the invariance of under orthogonal transformation intrinsically provides the shell-model TBRE more opportunity to restore part of SU(3) properties, i.e. $Q(2^+_1)=\pm Q(2^+_2)$ correlations, even though these $Q$ correlations are insensitive to the SU(3) rotational spectrum as expected based on the experimental survey \cite{allmond}. On the other hand, IBM1 $Q(2^+_1)=\pm Q(2^+_2)$ correlations always favor low-lying rotational spectra, which indicates that the IBM is more strongly governed by the dynamic symmetry. The SU(3) group reduction rule also qualitatively explains why the Shell Model more obviously favors the $Q(2^+_1)=- Q(2^+_2)$ correlation compared with the IBM1.

Low-lying $Q$ correlations represent intrinsic nuclear collectivity, and thus are more sensitive to the wave-function detail than the spectrum. Therefore, the nuclear quadrupole collectivity may maintain in a far more deep level than the common realization based on the orderly spectral pattern.

\acknowledgements
The discussion with Prof. Y. M. Zhao and Prof. N. Yoshida is greatly appreciated. We also thank Dr. Z. Y. Xu for his careful proof reading. This work was supported by the National Natural Science Foundation of China under Grant No. 11305151.


\begin{thebibliography}{}
\bibitem{oes-1} A. Bohr, B. R. Mottelson, and D. Pines, Phys. Rev. {\bf 110}, 936 (1958).

\bibitem{oes-2} R. Rossignoli, N. Canosa, P. Ring, Phys. Rev. Lett. {\bf 80}, 1853 (1998).

\bibitem{oes-3} H. H\"{a}kkinen, J. Kolehmainen, M. Koskinen, P. O. Lipas, M. Manninen, Phys. Rev. Lett. {\bf 78}, 1034 (1997).

\bibitem{oes-4} W. Satu{\l}a, J. Dobaczewski, and W. Nazarewicz, Phys. Rev. Lett. {\bf 81}, 3599 (1998).

\bibitem{oes-5} T. Papenbrock, L. Kaplan, and G. F. Bertsch, Phys. Rev. B {\bf 65}, 235120 (2002).

\bibitem{haq} R. U. Haq, A. Pandey, and O. Bohigas, Phys. Rev. Lett.
{\bf 48}, 1086 (1982).

\bibitem{bohigas} O. Bohigas, R. U. Haq, and A. Pandey, in {\it Nuclear Data for Science and Technology}, edited by K. H. B\"{o}ckhoff (Reidel, Dordrecht, 1983), p. 809.

\bibitem{shriner} J. F. Shriner, Jr., G. E. Mitchell, and T. von Egidy, Z. Phys. A {\bf 338}, 309 (1991).

\bibitem{casten-1} R. F. Casten, N. V. Zamfir, and D. S. Brenner, Phys. Rev. Lett. {\bf 71}, 227 (1993).

\bibitem{casten-2} N. V. Zamfir, R. F. Casten, and D. S. Brenner, Phys. Rev. Lett. {\bf 72}, 3480 (1994).

\bibitem{rand-rev-1} V. K. B. Kota, Phys. Rep. {\bf 347}, 223 (2001).

\bibitem{rand-rev-2} V. Zelevinsky and A. Volya, Phys. Rep. {\bf 391}, 311 (2004).

\bibitem{rand-rev-3} Y. M. Zhao, A. Arima, and N. Yoshinaga, Phys. Rep. {\bf 400}, 1 (2004). 

\bibitem{rand-rev-4} H. Weidenm\"{u}eller and G. E. Mitchell, Rev. Mod. Phys. {\bf 81}, 539 (2009).

\bibitem{rand-book} V. K. B. Kota, {\it Embedded Random Matrix Ensembles in Quantum Physics} (Springer, Heidelberg, 2014).

\bibitem{johnson-prl} C. W. Johnson, G. F. Bertsch, and D. J. Dean, Phys. Rev. Lett. {\bf 80}, 2749 (1998).

\bibitem{johnson-prc} C. W. Johnson, G. F. Bertsch, D. J. Dean, and I. Talmi, Phys. Rev. C {\bf 61}, 014311 (1999).

\bibitem{bijker-prl} R. Bijker and A. Frank, Phys. Rev. Lett. {\bf 84}, 420 (2000).

\bibitem{bijker-prc} R. Bijker and A. Frank, Phys. Rev. C {\bf 62}, 014303 (2000).

\bibitem{horoi-be2} M. Horoi, B. A. Brown, and V. Zelevinsky, Phys. Rev. Lett. {\bf 87}, 062501 (2001).

\bibitem{horoi-a} M. Horoi and V. Zelevinsky, Phys. Rev. C {\bf 81}, 034306 (2010).

\bibitem{zhao-be2} Y. M. Zhao, S. Pittel, R. Bijker, A. Frank, and A. Arima, Phys. Rev. C {\bf 66}, 041301(R) (2002).

\bibitem{allmond} J. M. Allmond, Phys. Rev. C {\bf 88}, 041307 (2013).

\bibitem{elliott-su3} J. P. Elliott, Proc. Roy. Soc. A {\bf 245}, 128 (1958).

\bibitem{ibm} F. Iachello and A. Arima, {\it The interacting boson model} (Cambridge University, Cambridge, UK, 1987), and corresponding references therein. 

\bibitem{wigner} E. P. Wigner, Ann. Math. {\bf 67}, 325 (1958).

\bibitem{tbre-1} J. B. French and S. S. M. Wong, Phys. Lett. B {\bf 33}, 449 (1970).

\bibitem{tbre-3} O. Bohigas and J. Flores, Phys. Lett. B {\bf 34}, 261 (1971).

\bibitem{tbre-2} S. S. M. Wong and J. B. French, Nucl. Phys. A {\bf 198}, 188 (1972). 

\bibitem{bijker-mf} R. Bijker and A. Frank, Phys. Rev. C {\bf 64}, 061303 (2001).

\bibitem{rand-sen} Y. Lei, Z. Y. Xu, Y.M. Zhao, S. Pittel, and A. Arima, Phys. Rev. C {\bf 83} 024302 (2011).

\bibitem{elliott-q} J. P. Elliott, Proc. Roy. Soc. A {\bf245}, 562 (1958).

\bibitem{ahv} T. Tamura and T. Udagawa, Phys. Rev. {\bf 150}, 783 (1966)

\bibitem{ahv-q} J. W. Lightbody, Jr., S. Penner, and S. P. Fivozinsky, Phys. Rev. C {\bf 14}, 952 (1976).

\bibitem{chen-mf} P. Van Isacker and J. Q. Chen, Phys. Rev. C {\bf 24}, 684 (1981).
\end{thebibliography}
\end{document}